\documentclass[aps,prl,preprint,showpacs]{revtex4}
\usepackage{epsfig}
\usepackage{axodraw}
\usepackage{graphicx}

\newcommand{\gsim}{\lower.7ex\hbox{$\;\stackrel{\textstyle>}{\sim}\;$}}
\newcommand{\lsim}{\lower.7ex\hbox{$\;\stackrel{\textstyle<}{\sim}\;$}}

\begin{document}

\title{
Soft Leptogenesis in Higgs Triplet Model}

\author{Eung Jin Chun and Stefano Scopel}
\affiliation{Korea Institute for Advanced Study, Seoul 130-722,
Korea}


\begin{abstract}
We consider the minimal supersymmetric triplet seesaw model as the
origin of neutrino masses and mixing as well as of the baryon
asymmetry of the Universe, which is generated through soft
leptogenesis employing a CP violating phase and a resonant behavior in
the supersymmetry breaking sector.  We calculate the full
gauge--annihilation cross section for the Higgs triplets, including
all relevant supersymmetric intermediate and final states, as well as
coannihilations with the fermionic superpartners of the triplets.  We
find that these gauge annihilation processes strongly suppress the
resulting lepton asymmetry. As a consequence of this, successful
leptogenesis can occur only for a triplet mass at the TeV scale, where
the contribution of soft supersymmetry breaking terms enhances the CP
and lepton asymmetry. This opens up an interesting opportunity for
testing the model in future colliders.
\end{abstract}

\maketitle

Leptogenesis is an elegant way to generate the baryon asymmetry of the
Universe in connection with the origin of the observed neutrino masses
and mixing through the seesaw mechanism \cite{fy}.  One way of
understanding a tiny neutrino mass is to relate it with the small
vacuum expectation value of a Higgs triplet \cite{Tss} whose decay can
also induce the cosmological baryon asymmetry in the presence of at
least two Higgs triplets \cite{Tlepto} or a right-handed neutrino
\cite{hybrids} as required by the generation of non-trivial CP and
lepton asymmetry.  In the minimal supersymmetric version with one pair
of triplets, there is a new way of leptogenesis (called ``soft
leptogenesis'') in which CP phases in the soft terms can contribute to
generate the lepton asymmetry \cite{softL,kitano}.  Soft leptogenesis
in the minimal supersymmetric Higgs triplet model has been considered
first in Ref.~\cite{softT}.

In this paper, we revisit this last scenario to provide a careful
analysis on the quantities for the lepton and CP asymmetries and their
cosmological evolution by considering the full set of Boltzmann
equations including thermal masses and the temperature supersymmetry
breaking effects consistently.  We will also derive a set of simple
Boltzmann equations from the Maxwell-Boltzmann approximation taking
into account the difference between the Bose--Einstein and
Fermi--Dirac statistics, and show that they provide a fairly good
approximation to the full Boltzmann equations.

The most important effect included in our analysis is the
contribution of the gauge annihilation processes, which lead to a
significant reduction of the resulting lepton asymmetry for the
low Higgs triplet mass. The dynamics of such a system is analyzed
in Ref.~\cite{fry} for the case of the conventional baryogenesis
with heavy Higgs bosons in the $SU(5)$ unification scheme.  Our
analysis is extended to the lowest possible values of the Higgs
triplet mass where, as will be shown in the following, the
annihilation effect dominates over decays and inverse decays.
Another crucial ingredient of soft leptogenesis is the suppression
of the  asymmetry due to a small difference between boson and
fermion statistics at finite temperature.  We find that this
effect for low values of the triplet mass becomes subleading
compared to that due to soft supersymmetry--breaking terms. Let us
also note that the annihilation effect becomes irrelevant for a
triplet mass higher than about $10^{10}$ GeV \cite{hambye05}, for
which, however, the lepton asymmetry in the soft leptogenesis
scenario is also suppressed, as it is inversely proportional to
the triplet mass \cite{softL}.  As a result, we will conclude that
the required baryon asymmetry can be generated only at the
multi-TeV range of the Higgs triplet mass, and thus the model can
lead to distinct collider signatures through, in particular, the
production and decay of a doubly charged Higgs boson
\cite{gunion,chun03}. This opens up another interesting
possibility for generating the neutrino masses and mixing as well
as the cosmological baryon asymmetry at the TeV scale, which can
be tested in future colliders \cite{tev,apos}.

\smallskip

In the supersymmetric form of the Higgs triplet model \cite{anna}, one
needs to introduce a vector-like pair of
$\Delta=(\Delta^{++},\Delta^+,\Delta^0)$ and $\Delta^c=(\Delta^{c--},
\Delta^{c-},\Delta^{c0})$ with hypercharge $Y=1$ and $-1$, allowing for
the renormalizable superpotential as follows:
\begin{equation}
W= h LL \Delta + \lambda_1 H_1 H_1 \Delta  + \lambda_2 H_2 H_2
\Delta^c  + M \Delta \Delta^c
\end{equation}
where $h LL \Delta$ contains the neutrino mass term, $h \nu \nu
\Delta^0$. The soft supersymmetry breaking terms relevant for us
are
\begin{eqnarray}
{-\cal L}_{soft} &=& \left\{ hA_L LL \Delta + \lambda_1 A_1 H_1
H_1 \Delta \right.
\nonumber\\
&+&\! \left. \lambda_2 A_2  H_2 H_2 \Delta^c + BM\Delta \Delta^c
+ h.c. \right\} \nonumber\\
& +& m_\Delta^2  |\Delta|^2 + m^2_{\Delta^c} |\Delta^c|^2 .
\end{eqnarray}
Note that we have used the same capital letters to denote the
superfields as well as their scalar components.  We will consider
the universal boundary condition of soft masses; $A_L=A_1=A_2=A$
and $m_\Delta=m_{\Delta^c}=m_0$. 
In the limit  $M \gg m_0 ,A$, the Higgs triplet vacuum
expectation value $\langle \Delta^0 \rangle= \lambda_2 \langle
H_2^0 \rangle^2/M$ gives the neutrino mass
\begin{equation}
m_\nu = 2 h \lambda_2 {v_2^2 \over M} \,.
\end{equation}
The mass matrix of the scalar triplets is diagonalized by
\begin{eqnarray}
\Delta &=& {1\over\sqrt{2}} (\Delta_+ + \Delta_-)\nonumber\\
\bar{\Delta}^{c} &=& {1\over\sqrt{2}} (\Delta_+ -
\Delta_-)\nonumber
\end{eqnarray}
where $\Delta_{\pm}$ are the mass eigenstates  with the
mass-squared values, $M^2_{\pm}=M^2+ m_0^2 \pm BM$, and the
mass-squared difference, $\Delta M^2= 2 B M$.  In terms of the
mass eigenstates, the Lagrangian becomes
\begin{eqnarray}
{-\cal L} &=& {1\over \sqrt{2}} \Delta_{\pm} \big[ h\,
\tilde{L}\tilde{L} +
h(A_L \pm M)\, LL \nonumber\\
&& \lambda_1\, \tilde{H}_1 \tilde{H}_1 + \lambda_1 (A_1\pm M)\,
 H_1 H_1 \\
&&  \pm \lambda_2^*\, \bar{\tilde{H}}_2 \bar{\tilde{H}}_2 \pm
\lambda_2^*(A_2^*\pm M)\,\bar{H}_2 \bar{H}_2 \big] + h.c.
\nonumber
\end{eqnarray}
The heavy particles $\bar{\Delta}_{\pm}$ decay to the leptonic
final states, ${L}{L}, \bar{\tilde{L}}\bar{\tilde{L}}$, as well as
the Higgs final states, ${H}_1{H}_1,{\tilde{H}}_1{\tilde{H}}_1$
and $\bar{H}_2\bar{H}_2,\bar{\tilde{H}}_2\bar{\tilde{H}}_2$. Thus,
the out-of-equilibrium decay $\bar{\Delta}_{\pm}\to {L}{L},
\bar{\tilde{L}}\bar{\tilde{L}}$  can lead to lepton asymmetry of
the universe.

\smallskip

In order to discuss how to generate a lepton asymmetry in the
supersymmetric triplet seesaw model let us first consider the
general case of a charged particle $X$ ($\bar{X}$) decaying to a
final state $j$ ($\bar{j}$) and generating tiny CP asymmetric
number densities, $n_X-n_{\bar{X}}$ and $n_j-n_{\bar{j}}$.   The
relevant Boltzmann equations in the approximation of
Maxwell--Boltzmann distributions are
\begin{eqnarray} \label{boltzmann}
 {d Y_X \over d z} &=& - z K \left[ \gamma_D (Y_X-Y_X^{eq}) +
 \gamma_A {(Y_X^2-Y_X^{eq\,2})\over Y_X^{eq}}  \right] \nonumber\\
 {d Y_x \over d z} &=& - z K \gamma_D \left[ Y_x-
 \sum_k 2 B_k {Y_X^{eq}\over Y_k^{eq}} Y_k \right] \\
 {d Y_j \over d z} &=& 2 z K \gamma_D\left[ \epsilon_j  (Y_X-Y_X^{eq})
 + B_j ( Y_x - 2 {Y_X^{eq} \over Y_j^{eq} } Y_j ) \right] \nonumber
\end{eqnarray}
where $Y$'s are the number densities in unit of the entropy
density $s$ as defined by   $Y_X\equiv n_X/s \approx
n_{\bar{X}}/s$, $Y_x \equiv (n_X-n_{\bar{X}})/s$ and $Y_j\equiv
(n_j-n_{\bar{j}})/s$.  Here, the CP asymmetry $\epsilon_j$ in the
decay $X\to j$ is defined by
\begin{equation}
\epsilon_j \equiv {\Gamma(X\to j) -\Gamma(\bar{X}\to \bar{j} )
\over \Gamma_X }.
\end{equation}
In Eq.~(\ref{boltzmann}), $K\equiv \Gamma_X/H_1$ with the Hubble
parameter $H_1=1.66 \sqrt{g_*} M^2/m_{Pl}$ at the temperature $T=M$,
and $B_j$ is the branching ratio of the decay $X\to j$. For the
relativistic degrees of freedom in thermal equilibrium $g_*$, we will
use the Supersymmetric Standard Model value: $g_*=228.75$.

\begin{figure}
\includegraphics[width=0.65\textwidth,bb = 17 207 593 587]{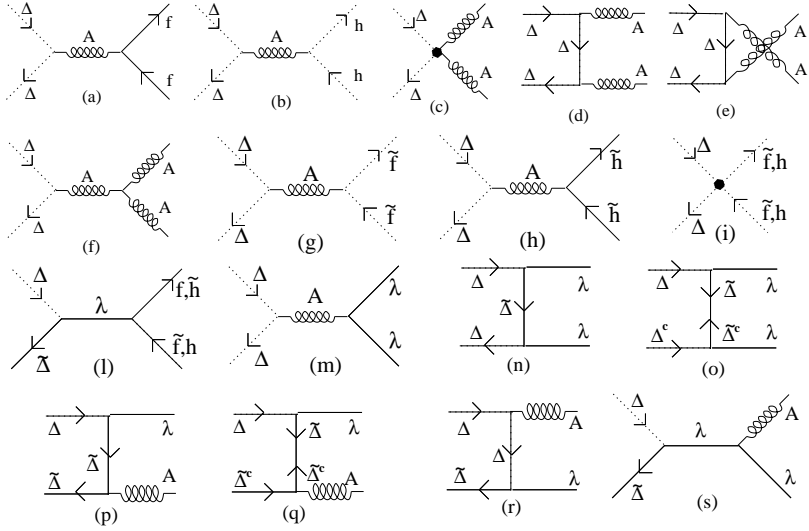}
\caption{\label{fig2_last} Diagrams contributing to the
gauge--annihilation amplitude of triplet particles.
$\tilde{\Delta}$, $\tilde{\Delta_c}$ represent the fermionic
partners of $\Delta$ and $\Delta_c$, respectively, while $A$
indicates a gauge boson, $\lambda$ a gaugino, $h$ a Higgs
particle, $\tilde{h}$ a higgsino, $f$ a fermion and $\tilde{f}$ a
sfermion. }
\end{figure}

The evolution of the $X$ abundance is determined by the decay and
inverse decay processes, as well as by the annihilation effect
described by the diagrams of FIG.~\ref{fig2_last}, and are
accounted for by the functions $\gamma_D$ and $\gamma_A$,
respectively.  Note that the triplets are charged under the Standard
Model gauge group and thus have nontrivial gauge annihilation
effect which turns out to be essential in determining the final
lepton asymmetry. Moreover, as a consequence of unitarity, the
relation $2 Y_x + \sum_j Y_j\equiv 0$ holds, so that one can drop
out the equation for $Y_x$, taking the replacement:
\begin{equation}
Y_x=-{1\over2} \sum_j Y_j
\end{equation}
in the last of Eqs.~(\ref{boltzmann}).
In our model, the heavy particle $X$ can be either of the six
charged particles; $X=\Delta^{++}_{\pm}, \Delta^{+}_{\pm}$ or
$\Delta^0_{\pm}$. Each of them follows the first Boltzmann
equation in Eq.~(\ref{boltzmann}) where $\gamma_D$ and $\gamma_A$
are given by
\begin{eqnarray}
\gamma_D &=& {K_1(z) \over K_2(z)} \label{eq:gamma_d}\\
\gamma_A &=& {\alpha_2^2 M \over  \pi K H_1}
 \int^\infty_1\!\! dt\, \frac{K_1(2zt)}{K_2(z)}\, t^2 \beta(t)\, \sigma(t)
\label{eq:sigmat_int}
\end{eqnarray}
with
\begin{eqnarray}
&&\sigma(t)=(14+11 t_w^4)(3+\beta^2)+(4+ 4 t_w^2+t_w^4)\left [
16+4(-3-\beta^2 + \frac{\beta^4+3}{2\beta}\ln
\frac{1+\beta}{1-\beta})\right ]\nonumber \\ &&+4 \left
[-3+\left(4-\beta^2+\frac{(\beta^2-1)(2-\beta^2)}{\beta}\ln\frac{1+\beta}{1-\beta}\right)
\right ],\label{eq:sigmat}
\end{eqnarray}
where $t_w\equiv\tan(\theta_W)$ with $\theta_W$ the Weinberg angle,
and $\beta(t)\equiv \sqrt{1-t^{-2}}$.  The function $\gamma_D$ is the
ratio of the modified Bessel functions of the first and second kind
which as usual takes into account the decay and inverse decay effects
in the Maxwell--Boltzmann limit.  The function $\gamma_A$ accounts for
the annihilation cross-section of a triplet component $X$ summing all
the annihilation processes; $X\bar{X}^\prime \to $ Standard Model
gauge bosons/gauginos and fermions/sfermions where $X^\prime$ is some
triplet component or its fermionic partner. The separate contribution
of each diagram in FIG.~\ref{fig2_last} is detailed in the
Appendix. As far as the Standard Model part is concerned, our result
agrees with that of Ref.~\cite{hambye05}, with one exception: the term
proportional to $t_w^2$, due to the mixed gauge boson ($W_3 B$) final
state in diagrams (c--f) of FIG.~\ref{fig2_last}, is missing in Ref.
\cite{hambye05}. However, this difference concerns a subdominant
contribution which is expected to have a negligible impact on
phenomenology.
\begin{figure}
  \includegraphics[width=0.65\textwidth]{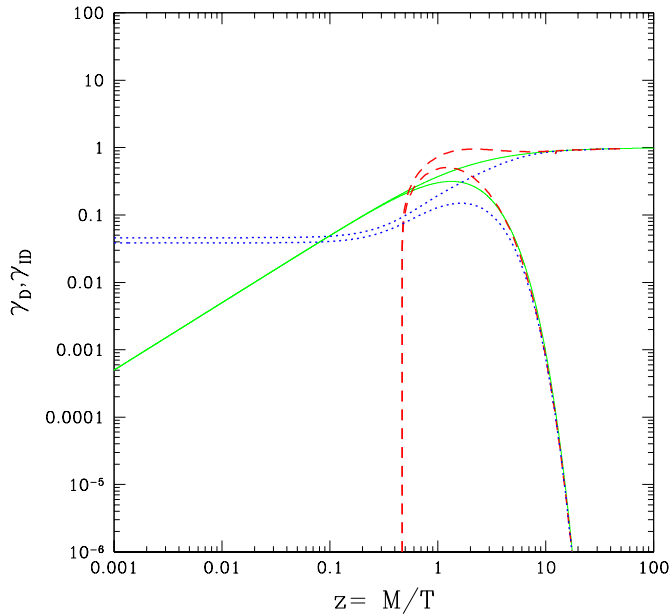}
\caption{\label{fig1_last} Decay and inverse-decay amplitudes
entering in the Boltzmann equations. The solid lines show the
decay amplitude $\gamma_D$ in the Maxwell--Bolzmann limit as given
by Eq.\ (\protect\ref{eq:gamma_d}), and the corresponding
inverse--decay amplitude $\gamma_{ID}$. The dotted and dashed
curves show the result of a full numerical evaluation of the same
amplitudes for fermionic and bosonic final states, respectively.
}
\end{figure}
The decay and inverse decay amplitudes in the Maxwell--Boltzmann limit
are plotted in FIG.~\ref{fig1_last}, along with a numerical evaluation of
the same quantities in the case of bosonic and fermionic final states,
where Bose--Einstein and Fermi--Dirac distributions, as well as
thermal masses, are included in the calculation. We use this latter
evaluation when we solve the full Boltzmann equations for the lepton
asymmetry numerically.  The last figure shows that the Boltzmann
approximation is well justified as expected for the region of our
relevance, $z>10$.
\begin{figure}
\includegraphics[width=0.65\textwidth]{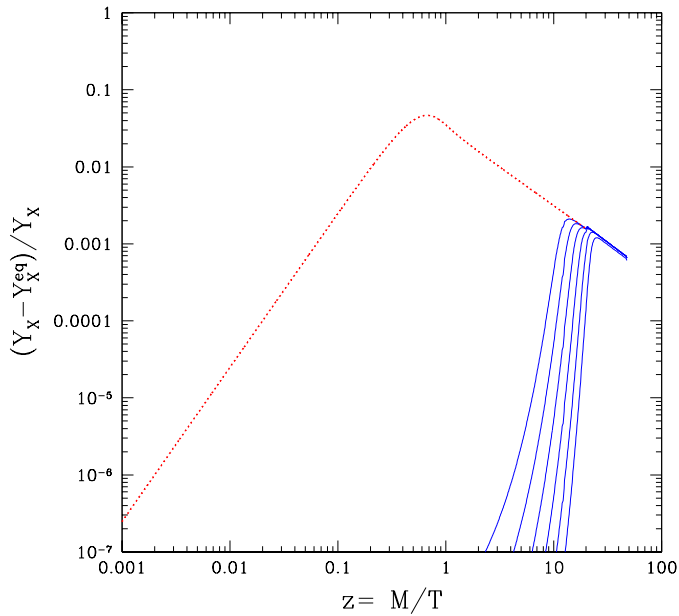}
\caption{\label{fig3_last} Fractional departure of the triplet
comoving density $Y_X$ from its equilibrium value $Y^{eq}_X$, as a
function of $z\equiv m/T$. The higher curve shows the result of a
calculation where the gauge annihilation effect is neglected,
while the lower ones show the same quantity including annihilation
for $Im(A)$=1 TeV and $\log_{10}(M/{\rm GeV})=$ 8,7,6,5,4,3 from
left to right.  All curves are evaluated in the Maxwell--Boltzmann
approximation.}
\end{figure}
Given $\gamma_D$ and $\gamma_A$, we can now analyze the thermal
evolution of $Y_X$ \cite{fry}. In FIG.~\ref{fig3_last}, we plot the
quantity $(Y_X-Y_X^{eq})/Y_X$, which quantifies the departure of
the triplet density from its equilibrium value. In particular, the
higher line shows the result when only the processes of decay and
inverse decay to light particles are included in the calculation.
As expected, since $K>>1$, $Y_X$ follows closely the equilibrium
density $Y_X^{eq}$ with a slight deviation of order $10^{-1}$.
However, annihilation is indeed important in our case and cannot
be neglected. This is shown in the same figure by the lower
curves, which represent the departure of the triplet density from
its equilibrium value when annihilation is included.  The
importance of annihilation can be understood in the following way.
The inverse decay freezes out at $z_f\approx 9$ for $K=32$ as $K
z_f^{5/2} e^{-z_f}=1$.  On the other hand, the thermal averages of
the annihilation and decay rate can be compared by considering the
following ratio \cite{fry}:
$${<\Gamma_A> \over <\Gamma_D>} (z_f)\simeq 2 {\alpha^2\over
\alpha_X} z_f^{-3/2}e^{-z_f} \approx {2\times 10^8\mbox{ GeV}\over
M },$$ where $\alpha_X = K H_1/M$.    Thus,  the annihilation
effect becomes negligible for $M \gsim 10^9$ GeV. But in our case
of soft leptogenesis, higher $M$ suppresses the lepton asymmetry
as $\tilde{\epsilon}_l \propto A/M$, so there is a tension between
these two effects, and lower values of $M$ turn out to be favored.
In FIG.~\ref{fig3_last}, one can see that, due to annihilation
which freezes out at $z\approx 20$, $Y_X$ follows more closely its
equilibrium density $Y_X^{eq}$ compared with the previous case,
with a deviation which is now of order $10^{-3}$. In particular,
this implies that the approximation
\begin{equation} \label{YXapp}
Y_X-Y_X^{eq} = {-Y_X^{eq
\prime}\over zK(\gamma_D+2\gamma_A)}
\end{equation}
is a good one, since decoupling occurs indeed at high
$z$. Nevertheless, in our numerical analysis, we solve the full
Boltzmann equations where the Bose--Einstein and Fermi--Dirac
distributions as well as thermal masses, are included properly.

\smallskip

To find out the cosmological lepton asymmetry by the decay of $X=
\Delta_{\pm}$, one needs to calculate $Y_j$ with the states $j=
LL$ and $\tilde{L}\tilde{L}$ and thus the corresponding CP
asymmetry:
\begin{equation}
\label{epsl}
 \epsilon_{L,\tilde{L}}
 \equiv {\Gamma(\bar{\Delta}_{\pm} \to LL, \tilde{L}\tilde{L})
- \Gamma( \Delta_{\pm}\to
\bar{L}\bar{L},\bar{\tilde{L}}\bar{\tilde{L}} ) \over \Gamma_{\pm}
}.
\end{equation}
Recall that one cannot rely on the the above Boltzmann equation
(\ref{boltzmann}) for the mechanism of soft leptogenesis in
the supersymmetric limit of $M\gg m_0, A, B$, as the CP asymmetries
in the bosonic and fermionic final states takes the opposite sign,
$\epsilon_L=-\epsilon_{\tilde{L}}$, so that the total asymmetry in
the lepton number density vanishes, $Y_l\equiv
Y_{L}+Y_{\tilde{L}}=0$.  A non-vanishing lepton asymmetry arises
after taking into account the supersymmetry breaking effect at
finite temperature \cite{softL}, namely the difference between the
bosonic and fermionic statistics given by the Bose--Einstein and
Fermi--Dirac distribution, respectively.  Such a thermal
supersymmetry breaking effect can be well accounted by a slight
modification of the last Boltzmann equation of
Eq.~(\ref{boltzmann}) resulting from the extension of the usual
Maxwell--Boltzmann approximation to the second order, as we will
show below.

\begin{figure}
\includegraphics[width=0.65\textwidth]{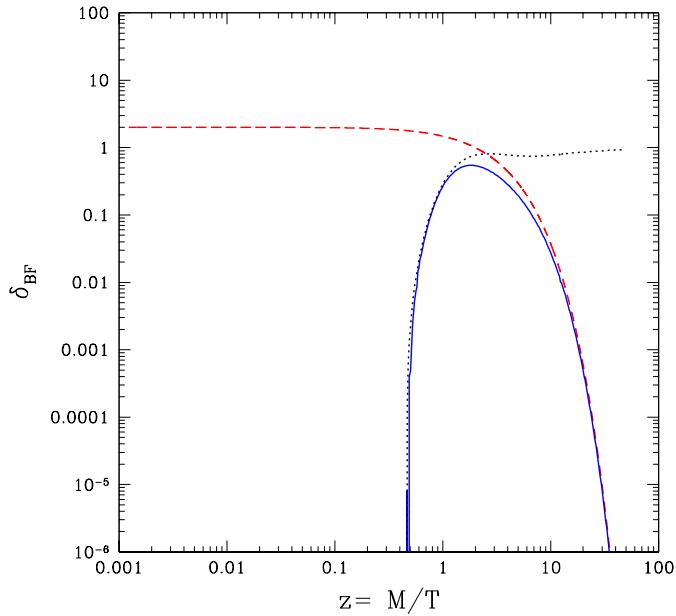}
\caption{\label{fig4_last}
The dashed curve shows the approximation to $\delta_{BF}(z)$ in
Eq.~(\protect\ref{eq:delta_bf}), while the solid line is the
result of a numerical evaluation of the same quantity, which
includes the effect of thermal masses and of Fermi--Dirac and
Bose--Einstein distributions. The dotted curve shows the result of
a numerical calculation of the thermal average $R_{H_2}
C_{\tilde{L}}$ which multiplies the soft supersymmetry breaking term
in Eq.~(\protect\ref{eq:lepton_asym}).}
\end{figure}

The complete form of the Boltzmann equation for the CP asymmetry
in the final state $j$ contains
\begin{eqnarray}\label{CPasym}
{s H_1\over z} {d Y_j \over d z} &\equiv& \int d\Pi_X d\Pi_{j_1}
d\Pi_{j_2} [
|A_j|^2 - |\bar{A}_{\bar{j}}|^2 ]\nonumber\\
&&[f_X(1\pm f_{j_1})(1\pm f_{j_2}) - f_{j_1} f_{j_2} (1+f_{X})]
+\cdots
\end{eqnarray}
where $d\Pi$'s are the phase space integration factors and  $A_j\,
(\bar{A}_{\bar{j}})$ is the amplitude of the decay $X\to j\,
(\bar{X}\to\bar{j})$.  The distribution functions $f_{j_i}$ at
thermal equilibrium  are $f_{B_i}=1/(e^{(E_i/T)}-1)$ or
$f_{F_i}=1/(e^{(E_i/T)}+1)$ for the bosonic or fermionic state
$j$.   Using the effective field-theory approach of resummed
propagators for unstable particles \cite{resonantL}, the effective
vertices of $\Delta_{+}$ ($\bar{\Delta}_+$) and the states $j$
($\bar{j}$) are
\begin{eqnarray}
S^j_+ &=& y^j_+ - y^j_- { i\Pi_{-+} \over \Delta M^2 + i \Pi_{--}
}
\nonumber\\
\bar{S}^{\bar{j}}_+ &=& y^{j*}_+ - y^{j*}_- { i\Pi_{-+}^* \over
\Delta M^2 + i \Pi_{--} }
\end{eqnarray}
where $\Delta M^2 = 2 BM$.  For $\Delta_-$, one takes the
interchange of $+ \leftrightarrow -$ and $\Delta M^2\to -\Delta
M^2$.  Here, $\Pi$'s are the absorptive part of two point
functions;
\begin{eqnarray} \label{Pis}
\Pi_{\pm\pm} &=& \sum_k {y^{k*}_{\pm} y^{k}_{\pm}\over 16\pi} R_k
\nonumber\\
\Pi_{\pm\mp} &=& \sum_k {y^{k*}_{\pm} y^{k}_{\mp}\over 16\pi} R_k
\,.
\end{eqnarray}
 Calculating $|A_j|^2-|\bar{A}_{\bar{j}}|^2
\propto |S^j_X|^2-|\bar{S}^{\bar{j}}_X|^2$, we get for
Eq.~(\ref{CPasym}),
\begin{eqnarray} \label{CPasym1}
&&[|A_j|^2 - |\bar{A}_{\bar{j}}|^2] (1\pm f_{j_1})(1\pm f_{j_2})
= \nonumber\\
&& -{4 \over 16\pi} \, \mbox{Im}(y^{j}_+ y^{j*}_- C_j \Pi^*_{-+})
\,
 {\Delta M^2 \over (\Delta M^2)^2 + \Pi_{--}^2} \,.
\end{eqnarray}
Here, $R_k$ include the thermal propagator effect in the cutting
rule \cite{Tloop} and $C_j$ are the thermal phase space factor of
the final states.
For the bosonic and fermionic states, 
we have
\begin{eqnarray}
R_B &=& \sqrt{1-4x_{B}} (1+f_{B_1}+f_{B_2} +2f_{B_1}f_{B_2})
\nonumber\\
R_F &=& (1-2x_{F})\sqrt{1-4x_{F}}(1-f_{F_1}-f_{F_2}
+2f_{F_1}f_{F_2})
\nonumber\\
C_B &=& \sqrt{1-4x_{B}} (1+f_{B_1})(1+f_{B_2})
\nonumber\\
C_F &=& (1-2x_{F})\sqrt{1-4x_{F}} (1-f_{F_1})(1-f_{F_2})
\end{eqnarray}
where $x_{B,F} = m^2_{B,F}(T)^2/T^2$ are the thermal masses of the
bosons or fermions.
%
Let us note in Eq.~(\ref{CPasym1}) that the
relation
\begin{eqnarray}
&&\sum_{j,k} \mbox{Im}(y^{j}_+ y^{j*}_-
C_j y^{k*}_+ y^{k}_- R_k)=
\nonumber\\
&& {1\over2}\sum_{j,k} \mbox{Im}(y^{j}_+ y^{j*}_-  y^{k*}_+
y^{k}_- ) (C_jR_k-C_kR_j)\propto \nonumber\\
&& {1\over2}\sum_{j,k} \mbox{Im}(y^{j}_+ y^{j*}_-  y^{k*}_+
y^{k}_- )(e^{E_1+E_2\over T} - e^{E_3+E_4\over T} ) \equiv 0
\nonumber
\end{eqnarray}
holds for any final states of $j_{1,2}$ and intermediate states in
the loop $k_{3,4}$.  The same is true for the second part of
Eq.~(\ref{CPasym}).  In fact, this is nothing but the unitarity
relation $\sum_j \Gamma(X\to j)= \sum_j \Gamma(\bar{X}\to\bar{j})$
from Eq.~(\ref{CPasym}).
Therefore, the lepton asymmetry in the integrand of the Boltzmann
equation (\ref{CPasym}) is found to be
\begin{eqnarray} \label{Lasym}
 &&~~ 2 |h|^2|\lambda_1|^2\, \big[ \mbox{Im}(A_L)M(|A_1|^2-M^2)
\nonumber\\
&&~~~~~~~~~~~~~~~ - \mbox{Im}(A_1)M(|A_L|^2-M^2)]C_L R_{H_1} \nonumber\\
&& + 2 |h|^2 |\lambda_2|^2\, \Big\{
 \big[ \mbox{Im}(A_L)M(|A_2|^2-M^2) \nonumber\\
&&~~~~~~~~~~~~~~~~~ + \mbox{Im}(A_2)M(|A_L|^2-M^2) \big] C_L R_{H_2}\nonumber\\
&&~~~~~ + \mbox{Im}(A_L)M M^2_\Delta C_L R_{\tilde{H}_2}   +
\mbox{Im}(A_1)M M^2_\Delta C_{H_2} R_{\tilde{L}} \Big\}\,.
\end{eqnarray}
Note that the terms proportional to $|h|^4$, which do not break
lepton number, disappear because of the previous relation of $C_L
R_{\tilde{L}}-C_{\tilde{L}} R_L=0$.  Thus, the asymmetry in
Eq.~(\ref{Lasym}) obviously contains only the mixed terms with
$h\lambda_{1,2}$, signaling a lepton number violation.
With the universality condition for the soft terms
($A=A_L=A_1=A_2$), we get a simple equation for the lepton
asymmetry as follows:
\begin{eqnarray}
&& 8 |h|^2 |\lambda_2|^2 \,\mbox{Im}(A) M^3 [\delta_{BF} +
\delta_{soft}
 ] \label{eq:lepton_asym} \\
\mbox{where}&&\delta_{BF} = {1\over2} [R_{H_2}(C_L-C_{\tilde{L}})+
C_{\tilde{L}}(R_{\tilde{H}_2}-R_{H_2})] \nonumber\\
\mbox{and} && \delta_{soft} = R_{H_2} C_{\tilde{L}} {m_0^2 + |A|^2
\over M^2}\,, \nonumber
\end{eqnarray}
putting $R=C=1$ in the denominator.
In the limit $M\gg m_0, |A|$,
we have
\begin{equation}
\Pi_{\pm\pm} = M \Gamma_{\pm} = {M^2\over 8\pi}
(|h|^2+|\lambda_1|^2 +|\lambda_2|^2)
\end{equation}
ignoring the small thermal effect and thus putting $R_k=1$.  One
thus finds that,  the quantity inside the integrand of
Eq.~(\ref{CPasym}) is proportional to
\begin{equation} \label{epspm}
{ 4 B \Gamma_\pm \over 4B^2 + \Gamma_\pm^2} {4 |h|^2 |\lambda_2|^2
\over (|h|^2 + |\lambda_1|^2 + |\lambda_2|^2)^2} {\mbox{Im}(A)
\over M}\, [\delta_{BF} + \delta_{soft}] \equiv
\tilde{\epsilon}_l\, [\delta_{BF} + \delta_{soft}]\,.
\end{equation}
Here, one has the approximation of
$\delta_{soft} = (m_0^2 + |A|^2)/M^2$ as can be seen in
FIG.~\ref{fig4_last}. One can also find similar expressions for the
Higgs--Higgsino final states. Recall that unitarity relation
enforces $\sum_j\epsilon^j_{\pm}=0$.   The supersymmetry breaking
effect $\delta_{BF}$ at finite temperature can now be encoded in
the Boltzmann equation with the Maxwell--Boltzmann approximation
by considering the expansion: $1/[\exp(E/T)\pm1]\approx
\exp(-E/T)[1\mp \exp(-E/T)]$.  After the phase space integration
in Eq.~(\ref{CPasym}), one obtains the simple modification of the
usual Boltzmann equation with the insertion of the $\delta_{BF}(z)$
function determined by
\begin{equation}
\delta_{BF}(z)\equiv 2\sqrt{2} {K_1(\sqrt{2}z) \over K_1(z)}
\label{eq:delta_bf}
\end{equation}
which gives a further suppression compared to the conventional
contribution with the Bessel Function $K_1(z)$.  The above
expression, which is monotonically decreasing in $z$, is valid for
$z\gg1$, and is compared in FIG.~\ref{fig4_last} to a numerical
calculation including the effect of thermal masses, which cause
$\delta_{BF}$ to vanish at small $z$.  The latter calculation of
$\delta_{BF}$ is obtained by numerically evaluating the thermal
average of the absorption part of the two--point function
$\Pi_{\pm\mp}$.
Concluding the above discussions, we find that the total lepton
asymmetry density $Y_l=Y_{L}+Y_{\tilde{L}}$ follows the
approximate Boltzmann equation:
\begin{equation} \label{boltzl}
 {d Y_l \over d z} = 2 g_\Delta z K \gamma_D
 \left[ \tilde{\epsilon}_l \delta(z)(Y_X-Y_X^{eq})
 + B_l ( Y_x - 2 {Y_X^{eq} \over Y_l^{eq} } Y_l ) \right]
\end{equation}
where $g_\Delta=6$ counts the total number of triplet components
generating the lepton asymmetry and $\delta(z)\equiv
\delta_{BF}(z)+\delta_{soft}$.   In the above equation, the number
$K=\Gamma_\pm/ H_1$ takes the minimal value of  $K=32$ for
$|h|=|\lambda_2| \gg |\lambda_1|$ as we have the relation
\cite{softT};
\begin{equation}
K=32 \,{|h|^2+|\lambda_2|^2 \over 2 |h| |\lambda_2|}
\left({|m_\nu| \over 0.05\mbox{ eV}}\right) \,.
\label{eq:K}
\end{equation}
As one goes away from the minimum value of $K$ with $|h|\neq
|\lambda_2|$, the quantity $\tilde{\epsilon}_l$ in
Eq.~(\ref{epspm}) gets suppressed. Furthermore, one realizes that
the resulting lepton asymmetry is maximized in case of
$B_{L,\tilde{L}}=B_{H_2,\tilde{H}_2} \gg B_{H_1, \tilde{H}_1}$
with $|h|=|\lambda_2|\gg |\lambda_1|$, in which case the Boltzmann
equation for the lepton asymmetry takes the simplest form of
\begin{equation} \label{Ylapp}
{d Y_l \over d z} = 2
g_\Delta z K \gamma_D \left[ \tilde{\epsilon}_l \delta(z)
 (Y_X-Y_X^{eq})
 - {Y_X^{eq} \over Y_l^{eq} } Y_l \right]
\label{eq:boltzmann}
\end{equation}
Let us now note that, taking the resonance condition
$B=\Gamma_{\pm}$, one finds the maximal value of
$\tilde{\epsilon}_l = {\mbox{Im}(A)\over M}$, which becomes order
one for $A \sim M \sim$ TeV.

\smallskip


\begin{figure}
  \includegraphics[width=0.65\textwidth]{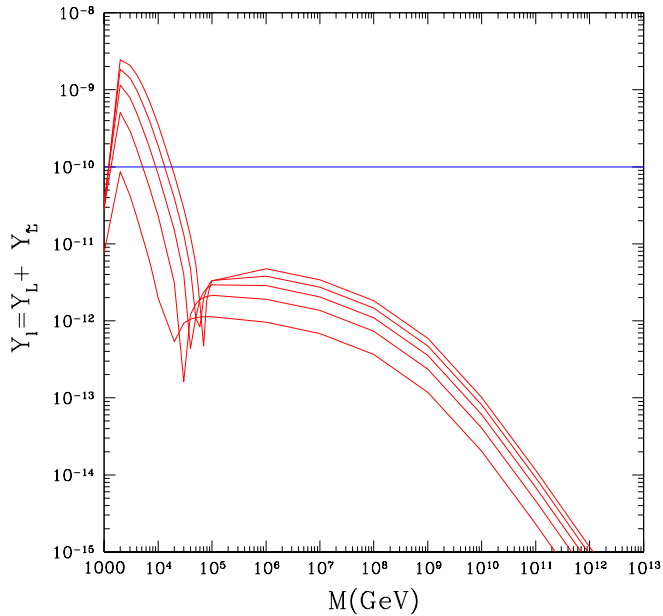}
\caption{\label{fig6_last} Final lepton asymmetry produced by
triplet decay as a function of $M$.  Different curves refer to
$\mbox{Im}(A)=1,2,3,4,5$ TeV from bottom to top.}
\end{figure}

It is now easy to find the approximate solution for $Y_l$ from
Eq.~(\ref{Ylapp}) with the insertion of $Y_X$ given in
Eq.~(\ref{YXapp}). Both are found to be a fairly good
approximation to the numerical solution of the full Boltzmann
equations, as expected from our previous discussions.  The results
of our numerical calculation are shown in FIG.~\ref{fig6_last}
where we plot the final lepton asymmetry as a function of the
triplet mass $M$ for $\mbox{Im}(A)$ from 1 to 5 TeV. When $|A|\sim
M$, one needs to recover the contributions of order $|A|/M$ which
were neglected, e.g., in Eqs.~(\ref{Pis}) and (\ref{epspm}).
Taking the parameter region, $m_0 < |A|=\mbox{Im}(A)$, we keep
those contributions in the numerical calculations for the curves
in FIG.~\ref{fig6_last}.  A remark is in order here.  For $M \gsim
10^{10}$ GeV, the annihilation effect becomes irrelevant and the
final asymmetry is determined by the decay and inverse decay
effects, i.e., by the value of $K$ only, which confirms the result
of Ref.~\cite{hambye05}.  One can see this feature in
FIG.~\ref{fig6_last}, where the lepton asymmetry as a function of
$M$ changes slope at about $M\sim 10^{10}$ GeV. Below this value
the annihilation effect sets in, and the final asymmetry is
strongly suppressed compared to the value one would obtain by
extrapolating the curve with the slope for $M\gsim 10^{10}$ GeV.

When the trilinear coupling is larger than the triplet mass,
$|A|/M \ge$ 1, besides an enhancement of the CP--violating term of
Eq.(\ref{epspm}), one could expect that the additional
contribution to the coupling of the triplet particles to scalar
final states enhances the total annihilation rate, increasing
substantially the value of the parameter $K$ compared to the
amount given by Eq.~(\ref{eq:K}), which is obtained in the limit
$|A|/M \ll 1$. As a consequence of this, the consequent additional
wash--out effect could in principle suppress the ensuing lepton
asymmetry. However, this is not the case due to the fact that, as
we have already discussed, the annihilation process freezes out
later than inverse decays, so the latter play almost no role in
the determination of the epoch when lepton asymmetry production
can start. Actually, this epoch starts when decays eventually
overcome annihilations, so a higher value of $K$ can slightly
anticipate it, leading so to a higher asymmetry instead than a
suppression, although this effect is quite mild. This is what we
observe in the numerical calculation shown in
FIG.~\ref{fig6_last}, where we have assumed as before
$|h|=|\lambda_2|\gg |\lambda_1|$ (in order to maximize the amount
of $CP$ violation given by Eq. (\ref{epspm})) and maximized the
$CP$--violating phase (i.e., we have assumed Re$(A)=0$).

We also remark that sphaleron interactions are kept in thermal
equilibrium even after the electroweak phase transition and freeze out
around $T_{sp}=M/z_{sp} \simeq 90$ GeV, so that only the lepton
asymmetry produced for $z<z_{sp}$ can be efficiently converted into a
baryon asymmetry \cite{apos}. As shown in FIG.~\ref{fig3_last}, due to
the gauge annihilation effect, the lepton asymmetry production is
delayed until $z=z_0\simeq 20$. For low values of $M$ ($\sim$ a few
TeV) one can have $z_{sp}<z_0$, which implies a suppression in the
final lepton asymmetry. This explains the fast rise at low values of
$M$ of all the curves in FIG.\ref{fig6_last}.  On the other hand, the
dips observed in the final asymmetry for $M\sim 20 |A|$ correspond to
the case when the two contributions $\delta_{BF}$ and $\delta_{soft}$
in Eq. (\ref{epspm}) are of the same order and cancel.  This dip
separates the two regions where $\delta_{BF}$ or $\delta_{soft}$
dominates in the determination of the final asymmetry. As shown in
FIG.~\ref{fig6_last}, the CP--violating contribution from the soft
supersymmetry breaking term $\delta_{soft}$ in
Eq.~(\ref{eq:lepton_asym}) can strongly enhance the final lepton
asymmetry at low values of $M$.  As a result, it is evident that the
required baryon asymmetry can be reached whenever $A$ and $M$ are in
the multi-TeV region.

Before concluding our work, let us remark some experimental
consequences of the model at future colliders. As shown above,
successful baryogenesis requires a TeV-scale triplet mass and
Yukawa couplings of the same order, $h\sim \lambda_2\sim
\sqrt{m_\nu M/v_2^2} \sim 10^{-6}$.  Thus, all  the low-energy
lepton flavor violating processes like $\mu\to e\gamma$ or $\mu
\to 3 e$ are highly suppressed \cite{chun03}.  On the other hand,
future accelerators have a potential to produce such Higgs
triplets, in particular, the peculiar doubly charged component
through the Drell-Yan processes \cite{gunion}. Then, various
features of the model can be checked by observing the branching
ratios of the triplet decay to lepton and Higgsino pairs, in
particular, $\Delta^{--} \to l^-_i l^-_j, \tilde{H}_2^-
\tilde{H}_2^-$, allowing also to study neutrino mass
patterns \cite{chun03}.

\smallskip

In conclusion, we have investigated baryogenesis assuming the
minimal supersymmetric Higgs triplet model as the origin of
neutrino masses and mixings. This model, with only one pair of
triplets, can provide a mechanism for soft leptogenesis employing
a CP violating phase and a resonant behavior in the supersymmetry
breaking sector. Our analysis shows that the original soft
leptogenesis, relying on the supersymmetry breaking effect
proportional to the small difference between boson and fermion
statistics at finite temperature cannot produce the right amount
of baryon asymmetry due to the gauge annihilation effect. In
particular, we have calculated the full gauge--annihilation cross
section including all the relevant supersymmetric intermediate and
final states, as well as coannihilations with the fermionic
superpartners of the triplets, finding that this effect strongly
suppresses the resulting lepton asymmetry.  On the other hand, the
contribution of soft supersymmetry breaking terms, particularly a
sizable value for the $Im(A)$ parameter, can enhance the lepton
asymmetry to provide successful leptogenesis if the triplet mass
is in the TeV range. In this case, the model predictions can be
tested in future colliders by searching for a very clean signal,
e.g., from the production and decay of doubly charged Higgs
bosons.

%

\appendix
\section[A]{Appendix}

In this appendix we give the detailed expression for the the
annihilation cross section shown in compact form in
Eq.(\ref{eq:sigmat}), and calculated from the diagrams (a)--(s) of
FIG. \ref{fig2_last}. In the following, masses of light particles
are neglected, while we  assume a common mass $M$ for the triplets
and their supersymmetric partners.
%
%
%
%
The reduced cross section introduced in Eq.~(\ref{eq:sigmat}) is
defined as:
$$\sigma(t)\equiv \frac{1}{2 g_2^4}\frac{1}{3} \sum\int
d\!\cos\theta\, |{\cal M}|^2,$$ in terms of the integrated squared
amplitude, averaged over the initial triplet state (hence the factor
1/3) and summed over the coannihilating particles, given by:
\begin{eqnarray}
\frac{1}{3}\sum \int d\!\cos\theta\, |{\cal M }|^2 &=&
  \frac{4}{3}\sum_a tr(T_a)^2 tr\left(\frac{\tau_a}{2}\right)^2\left[
    F_{(a)+(h)}+F_{(b)+(g)}+F_{(i)}+F_{(l)}
  \right]+\nonumber\\
&+&\frac{4}{3}\sum_{ab}\left\{tr(T_a^2 T_b^2)\left [
F_{(c)+(d)+(e)}+F_{(n)+(o)}+F_{(p)+(q)+(r)} \right ]+\right .
\nonumber\\&+& \left . 8 f_{abc} f_{abc} \left [
  F_{(f)}+F_{(m)}+F_{(s)}+F^{\prime}_{(d)}\right ] \right\}\,,
  \quad\mbox{where}
\label{eq:sigmat_detail}
\end{eqnarray}
\begin{eqnarray}
&&F_{(a)+(h)}=\frac{\beta^2}{3},\;\;F_{(b)+(g)}=\frac{\beta^2}{6},\;\;
F_{(i)}=\frac{1}{2},\;\;
F_{(l)}=1,\;\;F^{\prime}_{(d)}=-\frac{1}{2},\;\;F_{(s)}=-1\nonumber\\
&& F_{(c)+(d)+(e)}=2+2\left[
1-\beta^2+\frac{\beta^4-1}{2\beta}\ln\frac{1+\beta}{1-\beta}\right],
\;\;\;\;F_{(n)+(o)}=2+2\left[-2+\frac{1}{\beta}\ln\frac{1+\beta}{1-\beta}
\right ],\nonumber\\
&&F_{(p)+(q)+(r)}=4+2\left[-2+\frac{1}{\beta}\ln\frac{1+\beta}{1-\beta}
\right]\;\;\;\;F_{(f)}=\left
     [1-\frac{5}{6}\beta^2-\frac{(\beta^2-1)^2}{2\beta}\ln\frac{1+\beta}{1-\beta}\right ],
\nonumber\\
&& F_{(m)}=\frac{\beta^2}{3}+\left[1+\frac{\beta^2-1}{2\beta}\ln\frac{1+\beta}{1-\beta},
 \right].\nonumber
\label{eq:fterms}
\end{eqnarray}
\noindent In the last equation we have kept within squared parentheses
quantities that vanish for $\beta\rightarrow 0$, and the subscripts
refer to the contributing Feynman diagrams listed in
Fig.~\ref{fig2_last}. In Eq.~(\ref{eq:sigmat_detail}), $T_{a}$ and
$\tau_a/2$ are the SU(2)$\times $ U(1) group generators for the
triplet and doublet representations and $f_{abc}$ are the structure
constants.  Assuming the minimal supersymmetric Standard Model
particle content, the traces are given by:
\begin{eqnarray}
&&\frac{1}{3}\sum_a tr(T_a)^2 tr({\tau_a\over2})^2 =g_2^4(14+11
  t_w^4),\;\;\;\frac{1}{3}\sum_{ab}tr(T_a^2 T_b^2)=g_2^4(4+4
  t_w^2+t_w^4),
\label{eq:traces}
\end{eqnarray}
and $\sum_{ab}f_{abc} f_{abc}=6 g_2^4$.

Since annihilation decouples for $z \gg 1$, the integral in
Eq.~(\ref{eq:sigmat_int}) can be approximated by making use of the
following low-temperature expansion:
\begin{eqnarray}
&& \int^\infty_1\!\! dt\, \frac{K_1(2zt)}{[z K_2(z)]^2}\, t^2
\beta(t)\,
 \sigma(t)\simeq
\frac{1}{2 z^3}\left[b_0+\frac{b_1}{z}+... \right], \\
\mbox{where} &&b_0= 47+ 32 t_w^2 + \frac{49}{2} t_w^4 \quad
\mbox{and}\quad
b_1= -\frac{3}{2}\left (\frac{98}{3}+32 t_w^2+19
t_w^4 \right ). \label{eq:expansion}
\end{eqnarray}
Although for our results we used a numerical integration of
Eq.(\ref{eq:sigmat_int}), we have checked that the above
approximation leads to a good fit to the full numerical
calculation for $z\gsim$ 10, an interval that safely includes the
range of $z$ relevant for the present analysis. We finally notice
that the annihilation amplitude increases sizably in the
supersymmetric theory compared to the Standard Model case, in
significant excess of the factor {\cal O}(2) suggested by a
na\"{i}f expectation. In fact, the value of $b_0$ in
Eq.~(\ref{eq:expansion}) is almost 8 times larger than the
Standard Model value $b_0=6+8 t_W^2+2 t_W^4$ coming from the
diagrams (c)--(f). Such an enhancement is mainly due to a larger
number of available final states for the diagrams (i) and (l),
corresponding to the ``contact term'' for scalars and to
triplet--striplet annihilation to gauginos, respectively.


\end{document}